\documentclass[useAMS,usenatbib]{mn2e}

\usepackage{graphicx}
\usepackage{amsmath,amssymb}
\usepackage{times}

\newcommand{\Ms}{M_{\star}}
\newcommand{\Mcr}{M_{\rm CR}}
\newcommand{\Rs}{R_{\star}}
\newcommand{\Mbh}{M_{\bullet}}
\newcommand{\Mo}{M_{\odot}}

\newcommand{\peryr}{{\rm yr^{-1}}}
\newcommand{\ts}{t_{\star}}

\def\apgt{\ {\raise-.5ex\hbox{$\buildrel>\over\sim$}}\ } \def\aplt{\
{\raise-.5ex\hbox{$\buildrel<\over\sim$}}\ }

\bibpunct{\hspace{-2pt}}{}{}{a}{}{}

\title[Gravitational waves from remnants of ultraluminous X-ray
sources]{Gravitational waves from remnants of ultraluminous X-ray
sources}

\author[Clovis Hopman and Simon Portegies Zwart]{Clovis
Hopman$^{1}$\thanks{e-mail: clovis.hopman@weizmann.ac.il;
spz@science.uva.nl} and Simon Portegies
Zwart$^{2,3}$\footnotemark[1]\\ $^{1}$Center for Astrophysics, Faculty
of Physics, Weizmann Institute of Science, P.O. Box 26, Rehovot 76100,
Israel\\ $^{2}$Astronomical Institute ``Anton Pannekoek'', University
of Amsterdam, Kruislaan 403, Netherlands\\ $^3$Section Computational
Science, University of Amsterdam, Kruislaan 403, Netherlands}

\begin{document}

\date{Accepted ????. Received ?????; in original form ?????}

\pagerange{\pageref{firstpage}--\pageref{lastpage}} \pubyear{2004}

\maketitle

\label{firstpage}

\begin{abstract}
Ultraluminous X-ray sources (ULXs) with X-ray luminosities larger than
the Eddington luminosity of stellar mass objects may be powered by
intermediate mass black holes (IBHs) of masses $\Mbh\!\sim\!10^3\Mo$. If
IBHs form in young dense stellar clusters, they can be fed by Roche
lobe overflow from a tidally captured massive ($\Ms\!>\!10\Mo$) stellar
companion. After the donor leaves the main sequence it forms a compact
remnant, which spirals in due to gravitational wave (GW) emission. We
show that space based detectors such as the {\it Laser Interferometer
Space Antenna} are likely to detect several of these sources. GW
sources stemming from this scenario have small eccentricities which
give distinct GW signals.  Detection of such a GW signal will
unambiguously prove the existence of IBHs, and support the hypothesis
that some ULXs are powered by IBHs with captured companions.
\end{abstract}


\section{Introduction}
The counterpart of ultraluminous X-ray sources (ULXs), which have
X-ray luminosities $L_X$ larger than the Eddington luminosity of a
stellar mass object of mass $M$, $L_X>L_E=1.3\times10^{39}\, {\rm
erg\, s^{-1}}\,M/10\Mo$, are not known.  Most likely there is no
universal engine for ULXs. Some may be powered by anisotropic
radiation (e.g. King et al. \cite{K01}; Rappaport, Podsiadlowski \&
Pfahl \cite{RPP05}) or super Eddington luminosity (Begelman
\cite{B02}) from stellar mass black holes.

A third possibility is offered by the hypothesised existence of
intermediate mass black holes (IBHs; see Miller \& Colbert \cite{MC04}
for a review) of masses $10^3\,\Mo\lesssim\Mbh\lesssim10^5\Mo$, which could
radiate isotropically at a sub-Eddington rate to account for the
observed luminosities. There has been some observational evidence that
at least some ULXs are powered by IBHs. In some cases the
mass-temperature relation (Miller et al. \cite{MFMF03}; Miller, Fabian
\& Miller \cite{MFM04}) or quasi periodic oscillations (Fiorito \&
Titarchuk \cite{FT04}; Liu et al. \cite{L05}) indicate a high
accreting mass.

Numerical N-body and Monte Carlo simulations indicate that the stellar
collision rate in young dense stellar clusters during core collapse
can become very large, giving rise to a hierarchical merger. In that
case an object of thousands of solar masses may form (Portegies Zwart
et al. \cite{PMMH99}, \cite{PZea04}; G\"urkan, Freitag \& Rasio
\cite{GFR04}; Freitag, G\"urkan \& Rasio \cite{FGR05}; Freitag, Rasio,
\& Baumgardt \cite{FRB05}). The fate of such a massive star is
unclear, but it might lead to the formation of an IBH. This scenario
is supported by the fact that ULXs correlate positively with star
formation (Swartz et al. \cite{S04}; Liu, Bregman \& Irwin
\cite{LBI05}) and that some ULXs are associated with stellar clusters
(e.g. Zezas et al. \cite{ZFRM02}).

The young stars in the host cluster of the IBH have strong winds which
blow out the gas from the cluster, and there is insufficient free gas
available to power a ULX.  However, the IBH can acquire a companion
star by dynamical capture or tidal capture. Here we discuss the latter
possibility.  Hopman, Portegies Zwart \& Alexander (\cite{HPZA04})
showed that tidal capture of a main sequence star of mass $\Ms$ and
radius $\Rs$ can lead to circularization close to the tidal radius

\begin{eqnarray}\label{eq:rt}
r_t&=&\Rs\left(\frac{\Mbh}{\Ms}\right)^{1/3},
\end{eqnarray}
which is the distance from the IBH where the tidal forces equal the
forces which keep the star bound. As the star evolves it starts to
fill its Roche lobe causing it to lose mass to the IBH. This leads to
high X-ray luminosities, provided that the donor is sufficiently
massive ($\Ms\gtrsim10\Mo$, Hopman et al. \cite{HPZA04}; Portegies
Zwart, Dewi, \& Maccarone \cite{PZDM04}; Li
\cite{L04}). Interestingly, for some ULXs an optical counterpart has
been identified, indicating that these ULXs are binary systems (Liu,
Bregman, \& Seitzer \cite{LBS04}; Kuntz et al. \cite{K05}).

In this scenario a ULX may turn on for at most the life-time
$\ts\sim10\;{\rm Myr}$ of the captured star. In addition to strong
X-ray emission the star emits gravitational waves (GWs). Portegies
Zwart (\cite{PZ04}) discussed the possibility of observing X-rays and
GWs simultaneously. Only if the captured star is sufficiently light
($\Ms\lesssim2\Mo$), its tidal radius is small enough for it to emit
GWs in the LISA band during the mass transfer phase (Portegies Zwart
\cite{PZ04}).  However, in order to account for the high ULX
luminosities the companion star should be considerably more massive
than $2\,\Mo\,$ (Hopman et al. \cite{HPZA04}; Portegies Zwart et
al. \cite{PZDM04}; Li \cite{L04}).  In young and dense star clusters
mass segregation causes most massive stars to accumulate in the
cluster centre, and the combination of a high mass and large size make
these stars excellent candidates for tidal capture. When after time
$\ts$ the massive donor explodes it turns into a compact remnant (CR);
a neutron star (NS) or a stellar mass black hole (SBH). From that
moment the ULX, deprived of its source of gas, turns off.

Here we focus on the subsequent --post supernova-- evolution of the
(IBH, CR) binary. In-spiral of the CR into an IBH due to GW emission
results in a strong signal in frequencies measurable by space
detectors such as LISA, providing a wealth of information about the
system (see Miller \cite{M02} for a discussion). We show that LISA may
be able to observe such (IBH, CR) binaries.

\section{Stellar capture and binary evolution}\label{sec:cap}

When stars orbit IBHs on wide, but highly eccentric orbits, with
periapse close to the tidal radius of the IBH, $r_p\gtrsim r_t$,
orbital energy is invested in tidal distortions of the star, causing
it to spiral in. Hopman et al. (\cite{HPZA04}) showed that if the IBH
is less massive than $M_{\rm max}\approx10^5\Mo$, the star may be able
to cool down by radiating the excess energy efficiently, and survive
the strong tidal forces. Eventually it circularises near the IBH. We
first estimate the fraction $f_{\rm merge}$ of isolated binaries which
merge as a result of GW emission within the age of the Universe, while
accounting for angular momentum conservation during the mass transfer
phase and an isotropic velocity kick. We then discuss the consequences
of the gravitational interaction of cluster stars with the (IBH, CR)
binary.

\subsection{Evolution of an isolated IBH-star binary}

After tidal circularization near the IBH, the star may fill its Roche
lobe, possibly after a period of stellar evolution and
expansion. Roche lobe overflow from a $\Ms\gtrsim10\Mo$ donor can then
give rise to high luminosities for a time limited by the life-time
$t_\star$ of the star (Hopman et al. \cite{HPZA04}; Portegies Zwart et
al. \cite{PZDM04}; Li \cite{L04}).

The main-sequence star fills its Roche lobe at a distance $\sim2r_t$
before the terminal age main sequence. We assume that the entire
hydrogen envelope is transferred to the IBH, while the binary
conserves angular momentum. The mass of the stellar core of the donor
star is given by $M_c=0.08\;\Mo(\Ms/\Mo)^{1.4}$ (Iben, Tutukov \&
Yungelson \cite{ITY95}), while the mass of the hydrogen envelope is
$M_H=\Ms-M_c$. Angular momentum conservation implies that the binary
separation increases during the mass transfer phase.

The remaining helium star subsequently explodes in a super nova after
a time $\ts$, and forms a CR.  In this event a star of
$10\Mo<\Ms<20\Mo$ forms a NS of Chandrasekhar mass, $M_{\rm Ch} =
1.4\Mo$.  Stars more massive stars than $20\,\Mo\,$ collapse to SBHs;
we assume that the mass distribution is that found by Fyer \& Kalogera
(\cite{FK01}).

The semi-major axes and eccentricities of the CR are determined by the
post mass-transfer orbital elements while accounting for the mass lost
in the super nova and for the velocity kick, imparted to the CR at the
time of the explosion.  The kick velocity is taken in a random
direction with magnitude from the distribution of pulsars velocities,
which is well fitted by a double Gaussian distribution
\begin{eqnarray}\label{eq:vrt}
f(v)&=&4\pi v^2\Big[{w\over (2\pi \sigma_1^2)^{3/2}}\exp(-v^2/2\sigma_1^2)\nonumber\\
&& 
+{(1-w)\over (2\pi \sigma_2^2)^{3/2}}\exp(-v^2/2\sigma_2^2)
\Big],
\end{eqnarray}
where $w=0.4$, $\sigma_1=90\,{\rm km s^{-1}}$, and $\sigma_2=500\,{\rm
km s^{-1}}$ (Arzoumanian, Chernoff \& Cordes \cite{ACC02}). For SBHs
we adopt the same distribution, but with velocities smaller by a
factor $M_{\rm Ch}/M_{\rm SBH}$ (White \& van Paradijs
\cite{WVP96}; Gualandris et al. \cite{G05}). The velocity kick leads
to an increase of the eccentricity of binary and a change in its
energy; the kicks are generally insufficient to ionise the binary
systems (van den Heuvel et al. \cite{VDH00}).

The surviving binary loses orbital energy due to the emission of GWs.
A CR of mass $\Mcr$ on an orbit with semi-major axis $a$ and
eccentricity $e$ around an IBH of mass $\Mbh\gg\Mcr$ loses energy due
to the emission of GWs at rate of
\begin{eqnarray}\label{eq:DEdt}
\dot{E}_{\rm GW}={32\over5}{G^4\over c^5}{\Mbh^3\Mcr^2\over a^5}f(e),
\end{eqnarray}
with
\begin{eqnarray}\label{eq:fe}
f(e)=\frac{1+\frac{73}{24}e^{2}+\frac{37}{96}e^{4}}{(1-e^2)^{7/2}}.
\end{eqnarray}
If the orbit is circular $(e=0)$ spiral-in occurs on a time-scale of
(Peters \cite{Pe64}) 
\begin{eqnarray}\label{eq:t0}
t_{\rm merge}  &=&  {5\over256}
	            {c^5\over G^3}
	            {a^4\over\Mbh^2\Mcr}\nonumber\\
               &=&  3.5\,{\rm Gyr}\,
	            \left( {a \over {\rm AU} }\right)^4
	            \left({\Mbh\over3\times10^3\Mo}\right)^{-2}
	            \left({\Mcr\over10\Mo}\right)^{-1}.
\end{eqnarray}
For non-zero eccentricities the in-spiral time is shorter by a factor
$\sim(1-e)^{7/2}$.

We perform binary population synthesis with the above described
scenario to compute the fraction of high mass binaries that become
potential LISA sources.  The mass of the stellar companion was
selected from the initial mass function $dN/d\Ms\propto
\Ms^{-\alpha}$, where we assumed $\alpha=2$, consistent with the mass
function in the core of clusters in which a runaway merger occurred
(Portegies Zwart et al. \cite{PZ04}). We assume a minimal mass of
$10\Mo$, since lighter donors cannot account for ULX luminosities
(Hopman et al. \cite{HPZA04}; Portegies Zwart et al. \cite{PZ04}), and
a maximal donor mass of $100\Mo$.

LISA is likely to observe (IBH, CR) binaries in the phase before the
merger, rather than the merger event itself. However, the CRs spent
only a very short time in the LISA frequency band as compared to the
Hubble time (see \S\ref{sec:obs}). The question whether these sources
are detectable therefore depends on how many merge within a the Hubble
time.

\begin{figure}
\includegraphics[angle=270,scale=.37]{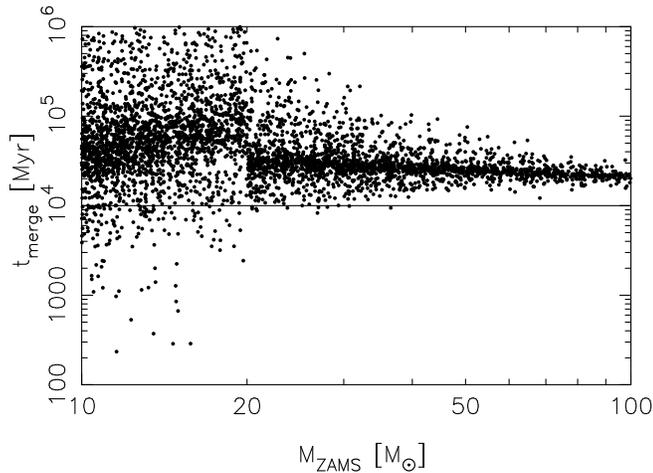}
\caption{Scatter diagram of the merger time $t_{\rm merge}$ as a
function of the zero-age main-sequence mass of the donor $M_{\rm
ZAMS}$. Events below the horizontal solid line merge within the age of
the Universe. For this image we generated 5000 binaries with a
$\alpha=2$ power-law initial mass function for the donor mass and we
selected an IBH mass of $\Mbh=3000 \,\Mo$. A small fraction (144 or
$2.9$\%) of the objects generated experienced merger within a Hubble
time.  The sudden jump at 20 $\,\Mo$ indicates the transition between
NS and SBH formation.\label{f1}}
\end{figure}

\begin{figure}
\includegraphics[angle=270,scale=.37]{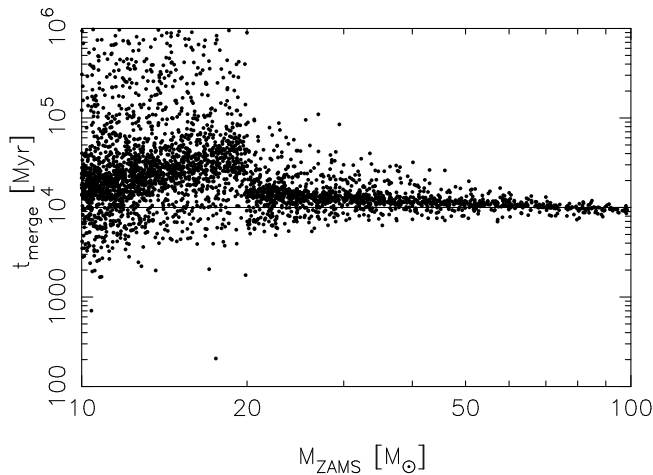} 
\caption{Same as figure (1), but with $\Mbh=10^4\Mo$. In this case
much more ($\sim15\%$) stars spiral in within the age of the Universe,
and in particular there is a significant contribution from SBHs. \label{f2}}
\end{figure}

\begin{figure}
\includegraphics[angle=270,scale=.37]{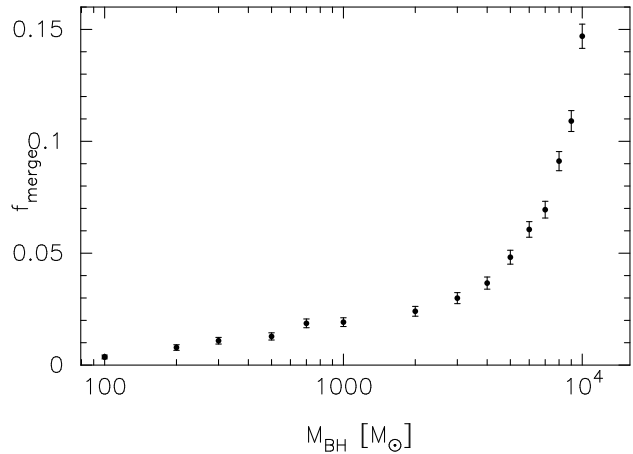}
\caption{Fraction $f_{\rm merge}$ of CRs which spiral in due to GW
emission within the age of the Universe as a function of the mass of
the IBH. The error bars are estimated by the 1-$\sigma$ Poissonian
uncertainties of the simulations.\label{f3}}
\end{figure}

In figures (\ref{f1}) and (\ref{f2}) we show the results for an IBH of
$\Mbh=3\times10^3\Mo$ and $\Mbh=10^4\Mo$, respectively. Only a
fraction of the star spirals in within a Hubble time. For
$\Mbh=3\times10^3\Mo$ nearly all objects which spiral in are NSs, with
only very few SBHs. For more massive IBHs, a significant fraction of
SBHs also spirals in fast enough, and the total fraction of merging
objects exceeds 10\% for $\Mbh \apgt 10^4\,\Mo$.  In
figure\,(\ref{f3}) we show $f_{\rm merge}$, the fraction of stars
which spiral in within the age of the Universe, as a function of
$\Mbh$.  After the velocity kick, the CRs have eccentricities up to
$e\lesssim0.9$. While this decreases $t_{\rm merge}$, these
eccentricities are much smaller than the $e\sim0.995$ eccentricities
found by Hopman \& Alexander (\cite{HA05}) for direct GW capture. By
the time the stars spiral in to orbital frequencies $\nu>10^{-4}\,$s,
which is in the LISA band, the orbits are close to circular.

\subsection{Interactions of the binary with cluster stars}\label{sec:scat}

The two-body relaxation time $t_r$ of star clusters which host a
runaway merger is short ($t_r\sim30\;{\rm Myr}$, Portegies Zwart \&
McMillan \cite{PM02}), and these evaporate on a time-scale of
$10^8$\,years if tidally limited. As long as the cluster has not yet
evaporated the (IBH, CR) binary has interactions with cluster
stars. Since the (IBH, CR) binary is ``hard'' (Heggie \cite{H75}),
these interactions tend to decrease the orbital separation between the
IBH and the CR (Miller \cite{M02}). In addition, scattering changes
the eccentricity of the binary. The angular momentum vector ${\bf J}$
of the binary performs a random walk, and its magnitude $J$ samples
angular momenta $0\lesssim J\lesssim J_m$ on the cluster's relaxation
time-scale (Alexander \& Hopman \cite{AH03}; Hopman \& Alexander
\cite{HA05}). Here $J_m$ is the maximum angular momentum of a binary
of given energy. When $J$ decreases the orbit becomes more eccentric,
and the in-spiral time decreases (see eq. [\ref{eq:t0}]).

It is not straightforward to quantify the effect of gravitational
two-body scattering, since the cluster is not in a steady state, and
in particular the relaxation time in the core can vary wildly. For
simplicity we therefore neglect the effect of scattering in the
following discussion on the in-spiral rate and the number of observable
LISA sources. We note, however, that scattering may significantly
increase the number of GW sources, so that the following LISA
detection rates should be regarded as lower limits.

\section{Observable gravitational waves from IBHs}\label{sec:obs}
The dimensionless strain of the GWs emitted at a frequency
$\nu=10^{-3}\nu_{-3}\,{\rm s^{-1}}$ from source at a distance
$d=d_{\rm Mpc}$ Mpc is

\begin{eqnarray}\label{eq:h}
h=3.5\times10^{-21}  \,\nu_{-3}^{2/3}\left({\Mbh\over3\times10^3\Mo}\right)^{2/3}{\Mcr\over10\Mo} d_{\rm Mpc}^{-1}
\end{eqnarray}
(e.g. Sigurdsson \& Rees \cite{SR97}).  LISA is sensitive to
frequencies in the range $10^{-4}\,{\rm Hz}\lesssim\nu\lesssim1$
Hz. At $\nu=10^{-3}$ Hz, LISA can detect sources with strains larger
than $h=\hat{h}_{-23}10^{-23}$, where $\hat{h}_{-23}\approx1$. This
estimate is based on a 1 yr observation with signal-to-noise ratio S/N
= 1 (see, e.g., http://www.srl.caltech.edu/lisa). In the following we
assume for concreteness a GW source with frequency $\nu=10^{-3}$ Hz;
application to other frequencies is straightforward. Sources can be
observed to distances up to
 
\begin{eqnarray}\label{eq:dmax}
d_{\rm max}\approx354 \,{\rm Mpc}\,\hat{h}_{-23}\nu_{-3}^{2/3}\left({\Mbh\over3\times10^3\Mo}\right)^{2/3}{\Mcr\over10\Mo}. 
\end{eqnarray}

We assume that only ULXs with luminosities $L_X>10^{40}\, {\rm erg
s^{-1}}$ contain IBHs (Portegies Zwart et al.
\cite{PZDM04}). Presently the average number of $>10^{40} \, {\rm erg
s^{-1}}$ ULXs per galaxy is $\sim0.1$ (Swartz et al. \cite{S04}). The
star formation rate dropped by an order of magnitude since $z\sim2$
(Madau, Pozzetti \& Dickinson \cite{MPD98}). ULXs correlate with star
formation; here we assume that the number of ULXs is proportional to
the star formation rate, in which case the number of luminous ULXs per
galaxy at an earlier time was $N_{\rm ULX}\sim1$. Since the ULX lives
for a time $\ts=10^7\,{\rm yr}\,t_7$, this yields an ULX formation
rate per galaxy of $10^{-7}\,\peryr N_{\rm ULX}t_7^{-1}$. As was
discussed in the previous section, only a small fraction $f_{\rm
merge}=0.1f_{-1}$ of the ULXs with $\Mbh<10^4\Mo$ leave behind a
remnant binary which spirals in within a Hubble time. The rate at
which observable GW sources are produced per relic ULX is then

\begin{eqnarray}\label{eq:gam}
\Gamma_{\rm GW}= 10^{-8}\,\peryr f_{-1}N_{\rm ULX}t_7^{-1}.
\end{eqnarray}

At the point where the period $P=2\pi a^{3/2}/(G\Mbh)^{1/2}$ equals
$10^{3}\,{\rm s}$, the in-spiral time is
\begin{eqnarray}\label{eq:tL}
t_L=155\,\,{\rm yr}\,\left({\Mbh\over3\times10^3\Mo}\right)^{-2/3}\left({\Mcr\over10\Mo}\right)^{-1},
\end{eqnarray}
where we assumed that the orbit has circularised when the frequency is
this high. We thus find that the mean number of sources emitting GWs
with frequencies of $\nu\sim10^{-3}\,{\rm s^{-1}}$ per galaxy is
$\mathcal{N}_{1}=\Gamma_{\rm GW}t_L$, or

\begin{eqnarray}\label{eq:Ngal}
\mathcal{N}_{1}=10^{-6}f_{-1}t_7^{-1}N_{\rm ULX}\left({\Mbh\over3\times10^3\Mo}\right)^{-2/3}\left({\Mcr\over10\Mo}\right)^{-1}.
\end{eqnarray}

The local galaxy density is estimated to be $n_{\rm
gal}\approx3\times10^{-2}\,{\rm Mpc^{-3}}$ (Marinoni et
al. \cite{M99}). If the maximal distance at which the GW source can be
observed is $d_{\rm max}$, the number of LISA sources in the sky at
any moment is given by $\mathcal{N}_{\rm tot}=\mathcal{N}_{1} n_{\rm
gal} 4\pi d_{\rm max}^3/3 $, or, using equation (\ref{eq:dmax}),

\begin{eqnarray}\label{eq:Ntot}
\mathcal{N}_{\rm tot} = 8.6 f_{-1}\hat{h}_{-23}^3t_7^{-1}N_{\rm ULX}\left({\Mbh\over3\times10^3\Mo}\right)^{4/3}\left({\Mcr\over10\Mo}\right)^{2}.
\end{eqnarray}

\section{Discussion}\label{sec:disc}

For IBHs of $\Mbh\lesssim3\times10^3\Mo$, $f_{-1}\sim0.3$ and almost
no SBHs spiral in (figure [\ref{f1}]). The main contribution to LISA
sources comes from NSs, i.e. $\Mcr/10\Mo\sim0.14$.We estimate that the
merger rate for NSs with IBHs is too low to be likely to be seen by
LISA. However, in this estimate we ignored the effect of three-body
scattering on the binary orbit. Though hard to quantise, its result
may be a marginal detection rate for NSs. We also note that we
estimated that SBHs have a velocity kick distribution similar to that
of NSs, but with velocities smaller by the mass ratio $M_{\rm
Ch}/M_{\rm SBH}$ of the objects. This estimate is rather uncertain,
and may be too conservative. Larger kick velocities cause more SBHs to
merge within a Hubble time, in which case the contribution of
$\Mbh\lesssim3\times10^3\Mo$ IBHs to LISA would increase.

For IBHs of $\Mbh \gtrsim 3\times 10^3\Mo$ we find that $f_{-1} \sim 1$
and for SBHs $t_{\rm merge}$ decreases to well within a Hubble
time. In this case stellar mass black holes $(M_{\rm CR}/10\Mo\sim1)$ give
the most promising GW sources, with a detection rate of about 10 per
year, in the case that three-body scattering is ignored.  Including
three-body scattering in this estimate may boost the detection rate
with a sizable fraction.

Our estimate in equation (\ref{eq:Ntot}) for the number of sources
that LISA can detect is conservative. First, in \S\ref{sec:obs} the
(IBH, CR) binary is considered to be isolated. Indeed the host cluster
eventually evaporates, but this is preceded by a phase during which
the binary interacts with other cluster stars
(\S\ref{sec:scat}). These interactions tend to harden the binary, and
change its eccentricity. As a result $t_{\rm merge}$ decreases, and
thus the number of potential LISA sources increases. It is not
implausible that nearly all (IBH, CR) binaries merge within a Hubble
time, in which case $f_{-1}=10$. Second, the life-time of ULXs is
probably significantly shorter than the main sequence life-time $\ts$
of the star. A more realistic assumption would be that ULX
luminosities are only achieved when the donor is near the terminal age
main sequence, in which case $t_7\approx0.1$, boosting the predicted
detection rate for LISA with an order of magnitude. In conclusion, the
number of observable GW sources could easily be orders of magnitude
larger than the expression (\ref{eq:Ntot}) indicates.

Previous studies of the number of potential LISA sources from binaries
with an extremely small mass ratio focused mainly on cases in which
the orbital energy is dissipated by the GWs themselves (e.g. Hils \&
Bender \cite{HB95}; Sigurdsson \& Rees \cite{SR97}; Ivanov
\cite{IV02}; Miller \cite{M02}; Freitag \cite{FR01}, \cite{FR03};
Alexander \& Hopman \cite{AH03}; Hopman \& Alexander \cite{HA05}). In
that case the event rate is on the order of a few per Gyr, and the
orbits are highly eccentric, with typical eccentricities as large as
$e\sim0.995$ for $\nu=10^{-4}\,{\rm s^{-1}}$ near an IBH of
$\Mbh=10^3\Mo$ (Hopman \& Alexander \cite{HA05}). In that case stars
spiral in very quickly, and emit GWs in the LISA band only for a short
time, in contrast to the here discussed tidal capture scenario.

GW sources originating from tidal capture sources have nearly circular
orbits when they enter the LISA frequency band, leading to a
distinctly different signal than the highly eccentric sources
originating from direct capture (e.g. Barack \& Cutler \cite{BC04};
Wen \& Gair \cite{WG05}). LISA can determine both the mass of the IBH
and the eccentricity of the orbit. Detection of GWs from an IBH will
give proof of the existence of these objects. If the signal stems from
an orbit with low eccentricity, this supports the scenario that ULXs
are accreting IBHs in binary systems.

\section*{Acknowledgments}
We thank T. Alexander for discussions. We are grateful to the Dutch
Royal Academy of Arts and Sciences (KNAW), The Dutch Organization for
Scientific Research (NWO) and the Dutch Advanced School for Astronomy
(NOVA).

\bsp

\label{lastpage}

\end{document}